# On the Efficiency of Ethics as a Governing Tool for Artificial Intelligence
## A critical comparison between AI Ethics and Bioethics

Pontifical Catholic University of Rio Grande do Sul

**Nicholas Kluge Corrêa**
Graduate Program in Philosophy
Pontifical Catholic University of Rio Grande do Sul
nicholas@airespucrs.org

**Nythamar De Oliveira**
Graduate Program in Philosophy
Pontifical Catholic University of Rio Grande do Sul
nythamar.oliveira@pucrs.br

**Diogo Massmann**
Graduate Program in Philosophy
Pontifical Catholic University of Rio Grande do Sul
diogo.massmann@edu.pucrs.br

October 26, 2022

## Abstract

The 4th Industrial Revolution is the culmination of the digital age. Nowadays, technologies such as robotics, nanotechnology, genetics, and artificial intelligence promise to transform our world and the way we live. Artificial Intelligence Ethics and Safety is an emerging research field that has been gaining popularity in recent years. Several private, public and non-governmental organizations have published guidelines proposing ethical principles for regulating the use and development of autonomous intelligent systems. Meta-analyses of the AI Ethics research field point to convergence on certain principles that supposedly govern the AI industry. However, little is known about the effectiveness of this form of "*Ethics*." In this paper, we would like to conduct a critical analysis of the current state of AI Ethics and suggest that this form of governance based on principled ethical guidelines is not sufficient to norm the AI industry and its developers. We believe that drastic changes are necessary, both in the training processes of professionals in the fields related to the development of software and intelligent systems and in the increased regulation of these professionals and their industry. To this end, we suggest that law should benefit from recent contributions from bioethics, to make the contributions of AI ethics to governance explicit in legal terms.

## 1 Introduction

Intelligent autonomous systems are increasingly becoming part of our society. Tasks and work previously performed by us are delegated to systems embedded with artificial intelligence (AI). The so-called 4th Industrial Revolution is the culmination of the digital age, where technologies such as robotics, nanotechnology, genetics, and artificial intelligence promise to transform our world and how we live. And, currently, the most accessible and massively used of these technologies is AI.

When we talk about AI regulation, our problems start with the question "*what is AI?*" A conclusive definition of AI, as recommended by international agencies like the United Nations Educational, Scientific and Cultural Organization (Unesco),[1] is currently an impossibility. Such a definition would change periodically as new technologies and technical resources allow innovations in the AI concept itself.

However, if we postulate that human/biological intelligence is what enables the execution of tasks that involve visual perception, speech recognition, decision-making, and natural language processing (among many others), AI would be understood as a computational system that integrates models and algorithms capable of coping with the complexity of such tasks.

These models and algorithms are what people refer to when speaking about AI. A field that as its current paradigm presents methodologies like machine learning, deep learning, reinforcement learning, and genetic algorithms, among many others [34]. Taking a more philosophical/cognitive science approach, following John Searle [36], one can still make the distinction between *weak AI*[2] and *strong AI*[3]. Over time, the definition of "*strong AI*" has come to refer to terms like "*human-level AI*"[4] or "*Artificial General Intelligence*" (AGI). [5]

To illustrate the reach that such systems have, we can cite the examples below as some types of AI used by the Brazilian public authorities:

- The *National Telecommunications Agency* uses natural language models to identify standard consumer behavior;
- The *National Land Transport Agency* uses AI to predict the average daily flow of traffic on federal highways;
- *Bank of Brazil* uses natural language models (chatbots) for customer service, CNNs for facial recognition;
- *Federal Savings Bank* uses AI to predict fraudulent electronic transactions;
- The C*oordination for the Improvement of Higher Education Personnel* uses AI to confirm the authorship of publications and academic projects;
- The *Federal Police Department* uses AI for facial recognition and natural language models for risk prediction (e.g., fraud detection);
- *Brazilian Agricultural Research Company* uses AI to predict the best vegetable to be grown in a crop and image classification to detect diseases in crops;
- The *National Institute of Social Security* uses AI to predict irregularities in the issuance of social benefits;
- The *Federal Supreme Court* uses AI to categorize legal proceedings under general repercussions;
- The *Supreme Court of Justice* uses AI to perform automatic scans for each appeal presented to the court, cross-reference appeals with previous rulings (search for legal precedents), and generate recommendations for Ministers.

And there are many more cases of application of this type of technology, not only by the public sector but by all sectors that, in some way, interact with our society.

Given the transformative power of AI, it would be important for us to have a consensus on the norms and guidelines that govern such technologies. The field of AI ethics and safety and (AI Safety and AI Ethics) are emerging research areas that have gained popularity in recent years. Several private, public, and non-governmental organizations have published guidelines that propose ethical principles to improve the regulation of autonomous intelligent systems [33] [1] [18] [19].

In this article, we will focus on the normative challenges and problems inherent in the "*enforcement*" of AI Ethics. As much as this is a relatively new field of Applied Ethics, AI Ethics already has enough literature for meta-analyses of the field to have been carried out [22] [20] [12] [9]. One of the results that we can observe in these meta-analyses is that there is a convergence to a set of generic ethical principles (especially when AI Ethics is embroidered in a principled way): Accountability/Liability, Beneficence/Non-Maleficence, Children & Adolescents Rights, Dignity/Human Rights, Diversity/Inclusion/Pluralism/Accessibility, Freedom/Autonomy/Democratic Values/Technological Sovereignty, Human Formation/Education, Human-Centeredness/Alignment, Intellectual Property, Justice/Equity/Fairness/Non-discrimination, Labor Rights, Open source/Fair Competition/Cooperation, Privacy, Reliability/Safety/Security/Trustworthiness, Sustainability, Transparency/Explainability/Auditability, Truthfulness.

---

[1]Recommendation on the Ethics of Artificial Intelligence. Available in https://unesdoc.unesco.org/ark:/48223/pf0000380455.

[2]A type of artificial intelligence that is limited to a specific or narrow area.

[3]An artificial intelligence that constructs mental abilities, thought processes, and functions that are impersonated from the human brain.

[4]A type of artificial intelligence that can perform as well as a human being in any task.

[5]A type of artificial intelligence that can solve many problems (proficiently) and in a wide range of applications and domains.

Of these eleven principles, the most cited in the literature are [9]:

- *Transparency/Explainability/Auditability*: this set of principles supports the idea that the use and development of AI technologies should be transparent for all interested stakeholders. Transparency can be related to "*the transparency of an organization*" or "*the transparency of an algorithm*." This set of principles is also related to the idea that such information should be understandable to nonexperts, and when necessary, subject to be audited;
- *Reliability/Safety/Security/Trustworthiness:* this set of principles upholds the idea that AI technologies should be reliable, in the sense that their use can be verifiably attested as safe and robust, promoting user trust and better acceptance of AI technologies;
- *Justice/Equity/Fairness/Non-discrimination:* this set of principles upholds the idea of non-discrimination and bias mitigation (discriminatory algorithmic biases AI systems can be subject to). It defends the idea that, regardless of the different sensitive attributes that may characterize an individual, all should be treated "*fairly*."

However, one criticism that has been raised against the current state of AI Ethics is its apparent neglect of certain issues [20] [8] [21] [9]. The current underrepresentation of principles that try to bring light to issues related to, for example, labor rights, technological unemployment, lethal autonomous weapons, lack of diversity, and automated disinformation, may be an indication that ethics, by itself, may end up being an ineffective tool to solve such problems.

For come, AI Ethics, in its current state, is "*useless*" [29] and "*tothless*" [31].

After all:

- *What would be the real normative power of ethical guidelines for the AI industry and its developers?*
- *Would AI Ethics be an effective governance tool?*

These will be the questions that we will try to answer in this article. Furthermore, we will seek to draw a parallel between Bioethics/Medical Ethics and AI Ethics, to try to find a way out of this *principled* governance model.

## 2 Bioethics versus AI Ethics

Since the beginning of the 20th century, Ethics has traditionally been divided into three major domains: Metaethics, Normative Ethics, and Applied Ethics. Ethical theories seek to justify moral arguments/reasoning when dealing with common moral problems such as, "*why is killing morally wrong?*" or "*how should we lead a morally good life?*" [38].

But when talking about Applied Ethics, our focus goes to the problem of how to apply ethical theories to real decision-making processes ("*How much money should I invest in training new nurses, and how much in buying better equipment for my hospital?*").

We can use Bioethics as a "*success case*" of Applied Ethics, as ethical debates regarding issues of medical and health ethics (e.g., abortion, euthanasia, health care, stem cell research, cloning, eugenics, genetic research) have slowly gained projection and representation in our legislative sphere.[6] Remembering that while Bioethics encompasses Medical Ethics, the inverse is not.

This is an example of how ethical debates can, through the intervention of policie makers, become governance tools (e.g., laws). When using the concept of governance, we are referring to [24]:

> "*processes undertaken by a government, or any form of organization, on other organizations, formal or informal, in the form of laws, norms or power, to determine a previously established standard of conduct*".

With this definition in mind, we would like to raise the following question:

- *Are ethical guidelines (effective) governance tools for the AI industry?*

For Calo [5], ethical guidelines end up serving only as a marketing strategy, not mitigating risks or imposing real consequences when principles are transgressed:

> *Various efforts are underway within the industry, academia, and other organizations to address the ethics of AI. But these efforts probably cannot replace policy-making.*

---

[6]As an example, in most places, it is illegal to knowingly create a human clone, regardless of the purpose.

Criticisms like Calo's have become increasingly common in the literature. Floridi et al. [13] made a comparison between the ethical principles raised within AI Ethics and those defended by Bioethics/Medical Ethics. The authors showed that four classic principles are reinterpreted within AI Ethics(i.e., respect for human autonomy, harm prevention / non-maleficence, beneficence, and justice/equity).

However, we cannot assume that such a principled/foundationalist approach (within Medical Ethics and Bioethics itself) has (just like AI Ethics) no criticisms. We can mention Norman Daniels [10], who sought to transpose the Rawlsian idea of reflective balance to the context of Bioethics in his work "*Just Health Care*," where the author raises several problems of foundationalism inherent in principlism.

According to Daniels [10], a reflective approach (i.e., *reflective equilibrium*), in contrast to a principlist approach in bioethics, would provide a form of "*moral updating*", where new contexts cause a change in the interpretation of the very principles/fundamentals that underlie our ethics, and consequently, the laws that should regulate our society.Other similar critisisms can be found in the bioethical literature [37] [7].

Another "*worth-copying aspect*" of all research in Bioethics concerns the strong presence of multidisciplinary research at the interfaces of moral philosophy and legal sciences with medicine, biology, and health sciences in general. Thus, we increasingly notice that AI Ethics, as in the case of Bioethics, cannot do without the same type of multidisciplinary articulation that Bioethics promotes.

In addition, another example of multidisciplinary research in bioethics, according to the lapidary formulation of Adina Roskies [32], is the case of "*ethics of neuroscience*."[7] Like AI Ethics, this field also has to deal with the technological advances of another area (Neuroscience). As Walter Glannon [16] has aptly argued, the ability to map, intervene, and alter the mind's neural correlates certainly raises important ethical questions.

Now that we have drawn this parallel between Bioethics/Medical Ethics and AI Ethics, the question we raise is as follows:

- *Is a comparison between Bioethics/Medical Ethics, which in the West can be traced back to the times of Hippocrates and Socrates, with the much more recent field of AI Ethics, a fair comparison?*

For Mittelstadt [26], a comparison between AI Ethics and Medical Ethics can not be made in the current state of AI Ethics. According to the author, AI development (and all software development in general) lacks some critical points to make such a comparison a possibility. These points are:

- *Common goals and duties*: while the practice of Medicine is guided by a common goal, i.e., the restoration and maintenance of health, the AI industry does not have any kind of common goal. The AI industry's goals are dictated by market demand rather than a common moral goal.
- *Tradition*: Medicine has a long tradition of ethical principles that helped guide its regulation (e.g., Hippocratic Oath, Formula Comitis Archiatrorum, Declaration of Geneva, Declaration of Helsinki) and define what we mean by ethical conduct in the area of Medicine [2]. Meanwhile, AI Ethics still lacks a tradition of robust conduct.
- *Legal accountability*: there is a whole social apparatus that guarantees that if the practice of medicine is carried out in an unethical or unsafe manner, such practitioners must be penalized with real consequences. In addition, practitioners of Medicine have intense academic training where they must meet high-quality standards. After the end of their training, these professionals are still subject to losing their licenses and being regulated by vigilant ethics committees. This mechanism provides a way to link "misbehavior" with legal sanctions. In the case of software engineering, no license is required to practice the profession. Institutes and regulatory associations in the area, such as the IEEE (Institute of Electrical and Electronic Engineers) and the ACM (Association for Computing Machinery), do not have the power to affect the practice of these professionals [17]. Professions related to software development are not yet legally recognized as regulated professions in several countries, and there is no general licensing standard [40].

Mittelstadt [26] also argues that professions characterized by having a robust ethical framework built into their practice have:

- Specialized education and training to deal with ethical dilemmas;
- Commitment to public service and pro-social objectives;

---

[7]"*An interdisciplinary field focusing on ethical questions raised by our growing and constantly improving understanding of the brain and our ability to monitor and influence it.*"

- A high standard of care;
- Self-governance is instituted by organizations of professionals in the field that also license the practice of the profession.

However, as the profession of "*programmer*" or software developer is still not regulated in many countries (such as Brazil), we believe that it is not possible to say that professionals who develop intelligent autonomous systems meet this series of criteria. So, how can we translate the ethical principles raised by the AI Ethics literature (as Bioethics does) into robust governance mechanisms, if we do not have "*on whom*" to impose the laws/standards we create?

## 3 Ethics in Software Development

The development of intelligent systems is still largely a software engineering endeavor, where Machine Learning represents only a portion of this development [35].

With approximately 30 years of history, Ethics in Software Development is still a very recent field of research. Within this literature, we find authors such as Friedman [15], Bynum [4], and Davis and Nathan [11], who argue in favor of the idea of Value Sensitive Design.[8] However, these ideas (Human-Centeredness) are still little incorporated by the standard Software Engineering literature.

To support this proposal, we can cite a controlled study conducted by McNamara et al. [25]. The sole purpose of this study was to investigate whether ethical guidelines have a normative effect on software developers' decision-making. In their research, the authors evaluated 63 software engineering students and 105 professional software developers, analyzing whether the ethical guidelines of the ACM would have any influence on the participants' decisions when dealing with moral dilemmas related to software development. The result found by the authors was the following: "*Despite our stated objective, we found no evidence that the ACM code of ethics influences ethical decision-making*."

As much as there is a large number of published ethical guidelines, there is still little research focused on how we can implement ethical principles in the practice of Software Development [27]. Vakkuri et al. [39] argue that some of the reasons why the AI Ethics literature has received little attention from areas related to Software Development are:

- Research in AI Ethics is predominantly philosophical;
- The Software Development area does not see the point in addressing philosophical concerns;
- As such, AI ethics training is not part of the standard educational system (in, for example, Computer science).

In the study by Vakkuri et al. [39], the authors explored how ethical issues were addressed in startup-like environments.[9] The authors carried out a qualitative case study with three Finnish healthcare startups, using semi-structured interviews with their developers, supervisors, and managers.[10]

In the interviews, the authors sought to know what concrete measures, whether through programming tools, standards of conduct, software verification, or any other not mentioned companies were taking to make their products in line with ethical principles already well highlighted in the literature:

- Transparency/Explainability/Auditability (*How can we explain and understand the decision of an algorithm?*);
- Accountability/Liability (*What is the responsibility of the developers? In case of misuse or unforeseen side effects, how should developers handle it?*).

The authors of the study argue that while participants expressed (sometimes) concerns about ethical issues, there was no clear way how developers were addressing such concerns in the development of their products. Some excerpts from the interviews that pinpoint this "*ethical deficit*" we would like to cite are:

- Transparency/Explainability/Auditability: "*Developers recognize transparency as an objective, but it is not formally pursued.*"
- Accountability/Liability: "*Our responsibility is to keep the project on schedule*," and "*Developers have no plans to deal with unexpected system behavior resulting from, for example, Machine Learning or future expansion of the system's usage context.*"

---

[8]A way of designing technology taking into account human values during the process.
[9]A common breeding ground for AI applications.
[10]The healthcare area was chosen on the assumption that ethical considerations "*would receive greater attention*."

From these findings, we believe it is possible to argue that:

- There is a disregard on the part of certain software development companies concerning Professional Ethics;
- Certain companies are unaware of the importance of their role in mitigating the side effects of their products;
- There is a gap in the training of professionals in the area of Software Development;
- Software Development still lacks tools to apply ethical principles in its practical development.

Several other studies support the idea that ethical guidelines alone do not affect professionals' decision-making [3] [6] [23] [30]. And this idea resonates with several criticisms raised against the current state of AI Ethics. According to Jobin et al. [22]:

> *Private sector involvement in the field of AI ethics has been questioned for potentially using soft, high-profile policies as a way to turn a social problem into a technical one or to evade regulation altogether.*

Hagendorff [20]:

> *AI Ethics — or Ethics in general — has no mechanisms to reinforce its own normative claims.*

Rességuier and Rodrigues [31]:

> *Ethics has great powerful teeth. Unfortunately, we're barely using them in the AI Ethics — it's no wonder then that the AI Ethics is called toothless.*

And Corrêa and Oliveira:

> *Like critical theory, the Ethics of AI should focus on highlighting the neglected aspects of our society and its relationship to the tech industry, challenging its power structures so that the promise of a beneficial AI for all can be fulfilled. Not just as an ideal for the future of humanity, but for people today as well.*

The point we want to reinforce is that the social role of Ethics is not to be a soft version of the law, because that is not where ethical normativity finds its justifying power. And in a practical sense, what we need today is regulation, not just suggestions, opinions, and hunches. We argue that is a lot to be learned (and copied) from Bioethics to AI Ethics. Especially on how ethical concerns are grounded into robust governance frameworks that go "*beyond ethics*" to ensure "*ethical behavior*."

## 4 The need for Regulation

Some of the most recent meta-analyses performed in the field reveal that the bulk of normative documents published in the last two decades was produced by governmental institutions and private corporations (48%), followed by CSO/NGO (17%), non-profit organizations (16%), and academic institutions (12, 5%). It is indeed interesting to notice that the private sector has emited just as much "normative content" as the public sector.

Also, looking at this production from a historical perspective, we see, as Corrêa et al. puts [9], that the "*AI ethics boom*" can be tough as a significant increase in the production of normative documents (129 = 64, 5%) between the years 2017 and 2019.

However, when we look at the "Normative Strength" of such documents, as mesured by the authors of "Worldwide AI Ethics"[11], we see an evident lack of convergence to a more "*government-based*" form of regulation. The vast majority (98%) of documents analyzed only serve as "*soft laws*," i.e., guidelines that do not entail any form of a legal obligation, while only 4, 5% present stricter forms of regulation. Even more surprisingly, by filtering only documents produced by governmental institutions, the disproportion does not go away, with only 18, 7% of documents proposing legally binding forms of regulation.[12]

If not recognized by the authors of such documents, experts in the field are coming to agree that regulation may well be our only "*way out*."

---

[11]https://en.airespucrs.org/worldwide-ai-ethics.

[12]Countries like Canada, Germany, and the United Kingdom seem to be spearheading this trend. Countries that are known for having stronger "welfare" policies.

One of the most recent studies to address ethical issues related to notions of human rights in AI and Big Data governance was the SHERPA survey,[13], a Delphi study commissioned by the European Union. The study, through a questionnaire containing the questions below, obtained some of the following results:

- The five most urgent ethical issues in the field are "*lack of transparency*" (19), "*lack of privacy*" (17), "*algorithmic discrimination*" (17), "*loss of human decision-making*" (12), and "*control and malicious use of data*" (10);
- The most mentioned approach to address such issues was regulations through laws and legislation. Another proposal was the creation of "*Citizen juries*" as a means of encouraging dialogue between the AI tech industry and society at large.[14]

As we can see, there is a consensus that what we need is not new "*ethical guidelines*" but regulation. Controversially, the most common type of produced document by AI stakeholders is "*Guides*" instead of "*Bills*."

Once again, we would like to emphasize that the "*success*" of Bioethics can (mainly) be attributed to the fact that the debate generated by its theme was implemented in legislative processes, making professionals and entities working in the areas of health strongly regulated by laws, ethics councils, professional councils, etc.

However, we cannot say the same for the still "*digital western*" that governs the AI industry in an unruly way. Developers are not regulated or trained on a robust code of ethical conduct because we "*fear*" that this will "*slow down*" our technological advancement. But is acceleration the only way forward?

Ethics cannot be used as a regulatory tool for the AI industry. We need to move from principlism to regulation, and perhaps the first step towards this is the regulation of those who develop the technologies that promise to transform our society.

As some authors argue [28] [14], the justifications for the bills in a developing country like Brazil would be several, ranging from the emerging concern about human coexistence with autonomous/intelligent systems to issues of legal certainty and the protection of human rights, and the very sustainability of human life.

As a final critic against soft forms of regulation, on September 29, 2021, the Chamber of Deputies of the Federative Republic of Brazil approved Bill 21/2020, establishing foundations and principles for the development and application of artificial intelligence in the country. This bill represents the commendable initiative of (i) signaling to the world that Brazil is attentive to concerns pertinent to AI, (ii) providing greater legal certainty in the use of these technologies, and (iii) promoting the dissemination of AI in Brazil. However, would this be the legislative content of this bill?

Unfortunately, the principialism approach has barely gotten in Brazil. Something that may make the actual regulation of the AI industry a long ongoing project till a bill with "prohibitions and sanctions" becomes something to be considered. The critique of this bill is beyond the scope of this study, but a full technical report (pointing to many blind spots in the current formulation of Bill 21/2020, 5051/19, and 872/21) can be found following the link in the footnote.[15]

## 5 Conclusion

In this brief article, we sought to introduce the reader to the emerging field of research in AI Ethics, where one of its concerns is to establish the principles and values that should guide (and eventually regulate) the technological development of this kind of technology.

We also set out to show that this is nothing new, i.e., resorting to ethics to try to regulate the technological advance of a specific area of knowledge. Bioethics turns out to be an excellent example.

Currently (but not as if we took such issues as resolved), our moral concerns concerning issues such as cloning, genetic engineering, euthanasia, etc, have been transported from the ethical debate to the area of knowledge responsible for creating, interpreting, and defending the norms by which we agree to live in society, i.e., the Law.

Depending on where we live (as such norms vary from culture to culture), health professionals know the limits and duties of their work. And this is due to a series of factors before mentioned (i.e., common goals, tradition, and legal accountability). However, a parallel between Bioethics and AI Ethics can not be drawn without attending to some obvious disparities.

---

[13]Santiago, N. (2020). SHERPA Delphi Study—Round 1 Results [Project Deliverable]. SHERPA project. https://www.project-sherpa.eu/wp-content/uploads/2020/03/sherpa-delphi-study-round-1-summary-17.03.2020.docx.pdf.

[14]The democratization of AI governance [8].

[15]Technical Note on the Bills in Progress in the Federal Senate 2022 PL 21/20 - PL 5051/19 - PL 872/21. Availeble in https://en.airespucrs.org/nota-tecnica-aires.

Criticisms increasingly arise for the principled approach, as (only) abstract principles cannot regulate the AI industry. Ethical principles and human values can be interpreted in countless ways. The way Westerners understand the concept of "*privacy*" is perhaps a little different from the way Easterners understand the same principle. And there is nothing wrong with that.

One of the points we would like to stress (again) in this conclusion is that, in the current state of AI Ethics:

- Where their research is still predominantly philosophical, and tools that translate abstract principles into the practice of developing intelligent systems are still few;
- Where there is little buy-in from the general community to concerns raised by AI Ethics literature;
- Where developers are not regulated or supervised by any type of body or professional council;
- And where applied ethics are not (generally) part of the training curriculum for these actors;

It is not possible to say that AI Ethics, by itself, is capable of regulating the AI industry. Thus, it cannot be considered an "efficient" governance tool. Therefore, principialism should be only seen as a starting point and not the end destination.

If we do not wish AI Ethics to be anything more than a mere "*ethics washing*", we need, in Brazil and elsewhere, to find ways to turn our concerns into legislation capable of sanctioning and guiding the AI industry, just like Bioethics successfully did for many countries. And for that end, perhaps the first step we need to take is to formalize and regulate the performance of developers in this industry.

If we want to live in a future where AI is created based on laws rather than promises and votes of confidence, it may be necessary for the AI industry itself to finally go "*professional*" and institute the study of Ethics (applied to its context) as a fundamental part of its formation. Only then will the Law uphold the necessary tools, as is already done in the case of Bioethics and Medical Ethics, to regulate the use and development of artificial intelligence.